\documentclass[conference]{IEEEtran}
\IEEEoverridecommandlockouts
\usepackage{cite}
\usepackage{amsmath,amssymb,amsfonts}
\usepackage{mathtools, bigstrut}
\usepackage{multirow}
\usepackage{algorithm}
\usepackage{algpseudocode}
\usepackage{physics}
\usepackage{caption}
\usepackage{subfig}
\usepackage{tikz}
\usepackage{algpseudocode}
\usepackage{graphicx}
\usepackage{textcomp}
\usepackage{xcolor}

\usepackage{environ}
\makeatletter
\newsavebox{\measure@tikzpicture}
\NewEnviron{scaletikzpicturetowidth}[1]{%
  \def\tikz@width{#1}%
  \def\tikzscale{1}\begin{lrbox}{\measure@tikzpicture}%
  \BODY
  \end{lrbox}%
  \pgfmathparse{#1/\wd\measure@tikzpicture}%
  \edef\tikzscale{\pgfmathresult}%
  \BODY
}
\makeatother

\usepackage{hyperref}
\hypersetup{
  colorlinks   = true, %Colours links instead of ugly boxes
  urlcolor     = black, %Colour for external hyperlinks
  linkcolor    = blue, %Colour of internal links
  citecolor   = blue %Colour of citations
}
\def\BibTeX{{\rm B\kern-.05em{\sc i\kern-.025em b}\kern-.08em
    T\kern-.1667em\lower.7ex\hbox{E}\kern-.125emX}}
\begin{document}

% \title{Construction of non-CSS quantum codes using measurements on cluster states}
\title{Construction of non-CSS quantum codes using
measurements on cluster states}
%A complete procedure of constructing $[[n,1]]$ non-CSS quantum codes using measurements on cluster states}
\author{\IEEEauthorblockN{Swayangprabha Shaw\IEEEauthorrefmark{1},
 Harsh Gupta\IEEEauthorrefmark{2}, Shahid Mehraj Shah\IEEEauthorrefmark{3} and
 Ankur Raina\IEEEauthorrefmark{4}}
 \IEEEauthorblockA{\IEEEauthorrefmark{3}Department of ECE, National Institute of Technology Srinagar, India\\
 \IEEEauthorrefmark{1}Department of Physics, Indian Institute of Science Education and Research Bhopal, India \\
 \IEEEauthorrefmark{1}\IEEEauthorrefmark{2}\IEEEauthorrefmark{4}Department of EECS, Indian Institute of Science Education and Research Bhopal, India\\
 Email: \IEEEauthorrefmark{1}swayangprabha17@iiserb.ac.in,
 \IEEEauthorrefmark{2}harsh22@iiserb.ac.in,
 \IEEEauthorrefmark{3}shahidshah@nitsri.net,
 \IEEEauthorrefmark{4}ankur@iiserb.ac.in}}

\maketitle

\begin{abstract}
%\begin{normalsize}
The Measurement-based quantum computation provides an alternate model for quantum computation compared to the well-known gate-based model. 
It uses qubits prepared in a specific entangled state followed by single-qubit measurements. 
The stabilizers of cluster states are well defined because of their graph structure.
We exploit this graph structure extensively to design  non-CSS codes using measurement in a specific basis on the cluster state. 
% We aim to construct $[[n,1]]$ non-CSS code from a $(n+1)$ qubit cluster state.
The procedure is general and can be used specifically as an encoding technique to design any non-CSS codes with one logical qubit.
We show there exists a $(n+1)$ qubit cluster state which upon measurement gives the desired $[[n,1]]$ code.

% We have also shown in this paper to construct a cluster of $(n+1)$ qubits
% given the stabilizers of a $[[n,1]] $ code. 
% \textbf{\textcolor{red}{100 words we need currently 76 }}. 
%\end{normalsize}
 
\end{abstract}

% \begin{IEEEkeywords}
% cluster state, measurement based quantum computing, parity check matrix
% \end{IEEEkeywords}
\vspace{-0.2cm}
\section{Introduction }
\begin{figure}[h]
    \centering
    \begin{tikzpicture}[node distance={10mm}, thick, main/.style = {draw, circle, minimum size=0.4cm}]
\node[main] (1) {$\ket{+}$}; 
\node[main] (2) [ right of=1] {};
\node[main] (3) [right of = 2] {}; 
\node[main] (4) [right of= 3] {};
\node[main] (5) [below of = 1] {}; 
\node[main] (6) [below of= 2] {};
\node[main] (7) [below of = 3] {}; 
\node[main] (8) [below of = 4] {}; 
\node[main] (9) [below of = 5] {}; 
\node[main] (10) [below of = 6] {}; 
\node[main] (11) [below of = 7] {$V$}; 
\node[main] (12) [below of = 8] {}; 
% \node[main] (13) [below of = 9] {}; 
% \node[main] (14) [below of = 10] {$V$}; 
% \node[main] (15) [below of = 11] {}; 
% \node[main] (16) [below of = 12] {};
\draw (1) -- node[midway, above right, sloped ] {} (2);
\draw (2) -- node[midway, above right, sloped ] {} (3);
\draw (3) -- node[midway, above , sloped ] {$\mathrm{CZ}$} (4);
\draw (5) -- node[midway, above right, sloped ] {} (6);
\draw (6) -- node[midway, above right, sloped ] {} (7);
\draw (7) -- node[midway, above right, sloped ] {} (8);
\draw (9) -- node[midway, above right, sloped ] {} (10);
\draw (10) -- node[midway, above right, sloped ] {} (11);
\draw (11) -- node[midway, above, sloped ] {$E$} (12);
\draw (1) -- node[midway, above right, sloped ] {} (5);
\draw (5) -- node[midway, above right, sloped ] {} (9);
\draw (2) -- node[midway, above right, sloped ] {} (6);
\draw (6) -- node[midway, above right, sloped ] {} (10);
\draw (3) -- node[midway, above right, sloped ] {} (7);
\draw (7) -- node[midway, above right, sloped ] {} (11);
\draw (4) -- node[midway, above right, sloped ] {} (8);
\draw (8) -- node[midway, above right, sloped ] {} (12);
%-------------------------------------%
% \draw (9) -- node[midway, above right, sloped ] {} (13);
% \draw (10) -- node[midway, above right, sloped ] {} (14);
% \draw (11) -- node[midway, above right, sloped ] {} (15);
% \draw (12) -- node[midway, above right, sloped ] {} (16);
% \draw (13) -- node[midway, above right, sloped ] {} (14);
% \draw (14) -- node[midway, above right, sloped ] {} (15);
% \draw (15) -- node[midway, above right, sloped ] {} (16);

\end{tikzpicture}
 \caption{ Cluster as graph $G=(V,E)$ where $V$ is the set of vertices and $E$ is the set of edges.}   
 \label{fig:1}
\end{figure}
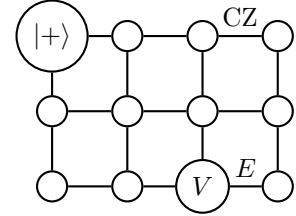

Quantum computing relies on the error-free operations of qubits. 
Inevitably, qubits suffer from noise and need to be preserved using Quantum error correcting codes (QECC). 
In the realm of quantum error correction, one of the elegant families of quantum codes is the Calderbank-Shor-Steane (CSS) codes named after their inventors \cite{nielsen2002quantum} whose structure motivates the construction of quantum codes using classical codes.
One of the important properties of CSS codes is that they consist of purely $X$ and $Z$ stabilizers \cite{gottesman1997stabilizer} which is particularly useful for the design of elegant decoders.
% Some popular examples of CSS codes include the Shor's code, which encodes one logical qubit into nine physical qubits \cite{shor1994algorithms}, and the Steane code, which has a higher coding rate and encodes one logical qubit into seven physical qubits \cite{duan1997preserving}. 
An interesting quantum code with an even better coding rate is the five-qubit code,  however that is a non-CSS code \cite{laflamme1996perfect}. 
This non-CSS code with least number of physical qubits is capable of correcting any single-qubit error. 
Motivation for the construction of such non-CSS codes stems from Measurement-based quantum computing (MBQC) proposed by Raussendorf \emph{et al.} \cite{raussendorf2003measurement}. 
This scheme is shown to be viable for universal quantum computing. 
It makes use of a resource state called \emph{cluster state} consisting of entangled qubits, initially prepared in the $\ket{+}$ state followed by single qubit measurement.
The model of computation is therefore also called cluster state computation \cite{jozsa2006introduction}.
%\textcolor{red}{[reference]}
The success of this approach is based on the creative utilisation of qubit measurements. %\textcolor{red}{(Rephrase this )}
A 2D grid cluster state is shown in Fig. \ref{fig:1}. 
This cluster state can be parameterized by mathematical structure given by $G=(V,E)$, where $G$ denotes the graph state associated with the cluster, $V$ indicates the set of vertices and $E$ represents the set of edges.
If one of the qubits in this cluster state is used as a message qubit then the resultant quantum state becomes a quantum error correcting code with one logical qubit.   

In this paper, we propose an algorithm showing explicit construction of encoding one logical qubit into $n$ physical qubits. 
Then we show how one can construct a $(n+1)$ qubit cluster associated with a $[n,1]$ code. 
We propose a general algorithm to construct a $(n+1)$ qubit cluster state from the parity check matrix of a given $[[n,1]]$ stabilizer code. Qubit measurement on $(n+1)$ qubit cluster gives the desired $[[n,1]]$ code, independent of the distance of the code. In this paper, we are not interested in increasing the distance of the code.

% We are giving only the encoding procedure for any $[[n,1]]$ non-CSS code.
The paper is organized as follows.
% Our method uses an initially prepared quantum state of $(n+1)$ qubits followed by  measurement of the strategically placed message qubit. 
% Though in the works of \cite{Schlingemann_2001}, \cite{https://doi.org/10.48550/arxiv.quant-ph/0111080} and \cite{2016NJPh...18j3052E} it has been shown that one can construct stabilizer codes using graph states, a compact algorithm regarding stabilizer evolution and construction of non-CSS codes using measurement is unique to this paper. 
% The construction is also verified via implementation on the IBM software kit called QISKit.
\vspace{-0.5cm}
%\vspace{-0.3cm}
We present the algorithm for the evolution of stabilizers for cluster states under measurement in Section \ref{sec:algo_stab} and show the procedure for construction of non-CSS codes in Section \ref{sec:algorithm1} with circuit simulations of the codes. We have shown our approach to construct the cluster corresponding to a given non-CSS code in Section \ref{sec:construction}. We conclude the paper in Section \ref{sec:conclude}. 

\vspace{-0.4cm}
\subsection{Notation}
% \begin{center}
\begin{tabular}{@{}ll@{}}
$X$, $Y$, $Z$ &  Measurement in $\sigma_x,\sigma_y,\sigma_z$ basis.\\
$\mathrm{CZ}$&$\mathrm{diag}(1,1,1,-1).$\\
$\mathrm{CZ}_{ij}$ & $\mathrm{CZ}$ Operation between $i^{\mathrm{th}}$ and $j^{\mathrm{th}}$ qubit.\\
			$\mathcal{N}(j)$ & Neighbourhood of site $j$.\\
            $S_j$ & $j^{\mathrm{th}}$ Stabilizer. \\
			$\mathcal{S}^{(i)}$ & Set consists of $<S_j>_{j=1}^{n},n$ is the number of qubits.\\
			$S_j^i$, $S_j^f$ & The initial and final $j^{\mathrm{th}}$ stabilizer respectively.\\
			$s$ & Parameter mapped to measurement outcome.\\
			$s_i$& $s$ associated with $i^{\mathrm{th}}$ qubit. \\
			$\mathcal{I}$& Set containing label of qubits need to measure. \\
			$\mathcal{M}$ & Set of measurable observables. \\
			$M_a$ & Elements in $\mathcal{M}$, where a $\in$ $\mathcal{I}$. \\
			$[S_j, M_a]$ & Commutation operation between $S_j$ and $M_a$.\\
			$\chi_1, \chi_2$ & $[S_j, X_a]$ and $[S_j, Z_a]$ respectively.\\
			$A,H$ & Adjacency and Parity check matrix respectively.
			\label{notation}
\end{tabular}
\section{Evolution of stabilizers}
\label{sec:algo_stab}
Consider the cluster state  shown in Fig. \ref{fig:1}, as a result of this graphical structure, the cluster state can be written as \cite{raussendorf2003measurement} :
\begin{align}
 \ket{G} = \displaystyle \prod_{(i,j) \in E} \mathrm{CZ}_{ij} \ket{+}^{\otimes n},
 \label{eq:1}
\end{align}
where $\mathrm{CZ}_{ij}$ is the entangling gate applied between every pair of nodes connected by an edge $(i,j)\in E$.
Here the nodes correspond to qubits prepared in $\ket{+}$ state. 
In this paper, we use nodes and qubits interchangeably.
An important feature of cluster states of $n$ qubits is that they have stabilizers of the form \cite{raussendorf2003measurement}:
\vspace{-0.25cm}
\begin{align}
\label{eq:2}
 S_j = X_j \displaystyle \prod_{k\in \mathcal{N}(j)} Z_k,\ j=1,2,\cdots, n.
 \end{align}
% where $\mathcal{N}(j)$ indicates the neighbors of node $j$. 
The unique nature of these stabilizers being not purely $X$ or $Z$ type has motivated us to construct non-CSS codes using the cluster states.

% The question we ask is how do the stabilizers evolve after certain set of measurements on the qubits in the cluster. 
Before going to non-CSS code construction, we first observe how the stabilizers evolve in response to a set of measurements on the cluster. 
% We consider an initial generator set of stabilizers $\mathcal{S}^{(i)}$ associated with an $n$ qubit cluster state. 
After performing $X$ or $Z$ measurement on qubits of different types of clusters, 
% we discard the measured qubit and remove the operator in stabilizer labelled same as the qubit.
we notice the final set of stabilizers contains only those stabilizers that commute with the measurement observables.
% Anti-commuting stabilizers are substituted with another stabilizer if they are two or more in number in the generator set. 
% The replacement of stabilizer is done by multiplying any two anti-commuting stabilizers from the set.
%\textcolor{red}{Needs editing}
Stabilizers having operator with same qubit index as measurement observables results in a phase factor $(-1)^s$ on the stabilizer where $s$ is the parameter mapped to measurement outcomes. 
The mapping of $s$ is such that $s=0$ when measurement outcome is $+1$ and $s=1$ when measurement outcome is $-1$.
This phase factor is helpful in adapting to the post-measurement operations on unmeasured qubits.
These observations lead us to the Algorithm \ref{alg:Algorith1} that gives the evolution of stabilizers upon measurement of qubits in the cluster.
\begin{algorithm}
\caption{Stabilizer Evolution, $\mathcal{S}^{(i)}\rightarrow\mathcal{S}^{{(m)}}$}\label{alg:Algorith1}
\begin{algorithmic}[1]
\State Initialize $\mathcal{S}^{(i)}=<S_{j}>_{j=1}^{n}$ where $S_{j}=X_{j} \displaystyle \prod_{k\in \mathcal{N}(j)} Z_{k}$
\State $\mathcal{M}=\{M_a, a \in \mathcal{I}\}$ 
\State $\mathcal{S}^{(k)}\subset \mathcal{S}^{(i)}\ \ \text{s.t} \ \{S_{l},\ M_a\}=0\ \forall \ S_{l}\in  \mathcal{S}^{(k)} $
\If{$|\mathcal{S}^{(k)}| \geq 2$}
\For {${S}_{q}\in \mathcal{S}^{(k)}$ }
\For{${S}_{p}\in \mathcal{S}^{(k)} \backslash \{{S}_{q}\}$ }
\State $S_{j}\leftarrow S_{p}S_{q}$ where $S_{j}\in S^{(i)}$ 
%\ \{S_{p}, S_{q}\}\in S^{(k)}$
\EndFor
\EndFor
\EndIf
\For {$M_{a} \in \mathcal{M}$}
%where  $I=\{1, 2, 3 \cdots n$\}
\State{$\chi_1:= [S_j,\ M_{a}=X_{a}]$}
\State{$\chi_2:= [S_j,\ M_{a}=Z_{a}]$}

\If {$\chi_1=0$ }
\If{$a=j$}
\State $(-1)^{s}S_{j}^{i}\leftarrow S_{j}^{i}$
\State $S_{j}^{f}=S_{j}^{i}$
\Else %{$I \neq j$}
\State {$S_{j}^{i}\leftarrow S_{j}^{i}$}
\State $S_{j}^{f}=S_{j}^{i}$
\EndIf
\ElsIf {$\chi_2=0$} 
%\implies a\neq j$
\If{$Z_a \in \displaystyle \prod_{k\in \mathcal{N}(j)} Z_{k}$}
\State {$(-1)^{s}S_{j}^{i}\leftarrow S_{j}^{i}$,}
\State $S_{j}^{f}=S_{j}^{i}$
\Else %{$a \neq j$}
\State {$S_{j}^{i}\leftarrow S_{j}^{i}$}
\State $S_{j}^{f}=S_{j}^{i}$
\EndIf 
\Else
\State {$S_j$ vanishes}
\EndIf
\EndFor
\State $\mathcal{S}^{(m)}=<S_{j}^{f}>_{j=1}^{n}$
\end{algorithmic}
\end{algorithm}

In this algorithm, the initial stabilizer generator group $\mathcal{S}^{(i)}$ contains $<S_j>_{j=1}^{n}$, whose expression is given in Eq. \ref{eq:2}.
%$\mathcal{I}$ denotes the labels of qubits on which measurements are performed. 
The number of measurements is equal to the cardinality of $\mathcal{I}$ ($|\mathcal{I}|$) defined in the $\textit{notation}$ \ref{notation}.
% Suppose we measure the observable  on the $a^{\mathrm{th}}$ qubit, then, $\mathcal{M}=\{M_a, a \in \mathcal{I}\}$ denotes the ordered set of measurement observables $M_a$.%, where $a \in \mathcal{I}$.
%For the observables in different basis, a new set $\mathcal{M}$ is proposed that contains $M_a$ observable acting on $a^{\mathrm{th}}$ qubit in the same order of numbers in $\mathcal{I}$.
% For example, in a measurement scheme where we need to measure the $1^{st}$, $3^{rd}$ , $6^{\mathrm{th}}$ qubits in $X, Z, X$ basis respectively, the set $\mathcal{I}$ can be written as $\{1,\ 3,\ 6\}$ and and the set of measurable observables $M=\{M_1,\ M_3,\ M_6\}$ can be written as $M=\{X_1,\ Z_3,\ X_6\}$.
Let consider $\mathcal{S}^{(k)}(\subset\mathcal{S}^{(i)})$ such that the elements in $\mathcal{S}^{(k)}$ anti-commutes with any of the measurement observable in the set $\mathcal{M}$ i.e., $\{S_l,\ M_a\}=0$, where $S_l\in\mathcal{S}^{(k)}$ and $M_a \in \{X,Y,Z\}$. 
If $|\mathcal{S}^{(k)}|$ is 2 or more, $S_p,S_q \in \mathcal{S}^{(k)}$ are replaced with $S_p S_q$.
The replacement stabilizer is appended to $\mathcal{S}^{(i)}$.
Then, to determine whether the measurement observable commutes with the updated $\mathcal{S}^{(i)}$ or not, a new variable $\chi_1$ and $\chi_2$ (defined in \textit{notation} \ref{notation}) is introduced.
For a commuting stabilizer ($\chi_1=0$ or $\chi_2=0$) with qubit index same as that of the measurement observable, the elements of $\mathcal{S}^{(i)}$ modifies as follows:
\begin{itemize}
    \item $M_a=X_a$ and $a=j$, the stabilizer modifies as $(-1)^s S_j$,
    \item $M_a=Z_a$ and $a=k$ where $k \in \mathcal{N}(j)$, $S_j$ then transforms as $(-1)^s S_j$.
\end{itemize}
The elements of $\mathcal{S}^{(i)}$ is appended then to a new generator set $\mathcal{S}^{(m)}$, which will be called as final set of stabilizers after measurement.

This algorithm is helpful in visualising the impact of $X$ and $Z$ measurement on cluster.
For example, in case of three-qubit cluster state which has the initial stabilizer group as $<X_1Z_2,\ X_2Z_1Z_3,\ X_3Z_2>$, the stabilizer generator set % and the resultant state is $X^{s_2}\ket{++}$ which leads us to the conclusion that

\begin{figure}[H]
    \centering
    \begin{tikzpicture}
     %  \draw [fill=black!6](-0.6,-1.5) .. controls (1,-1.45) and (2,-1.42) .. (3.8,-1.45) 
    \draw [rounded corners=5mm,fill=gray!8] (-0.05,-1.5) -- (3.6,-1.5) -- (3.6,-2.6) -- (-0.6,-2.6) -- (-0.59,-1.5) -- (-0.05,-1.5) ;
    
     \draw [black,thick,fill=green!30] (0,0) circle (0.4);
    \draw [black,thick,fill=green!30] (1.5,0) circle (0.4);
    \draw [black,thick,fill=green!30] (3,0) circle (0.4);
    \draw [black,thick,fill=green!30] (0,-2) circle (0.4);
    
    \draw [black,thick,fill=green!30] (3,-2) circle (0.4);
    
    % \draw [black,thick,fill=green!30] (1.8,-1.2) circle (0.4);

    \node at (0,0) {1};
    \node at (1.5,0) {2};
    \node at (3,0) {3};
    \node at (3,-2){3};
    \node at (0,-2) {1};
    % \node at (1.8,-1.15) {6};
    
    \draw [ thick] (0.4,0) -- (1.1,0);
    \draw [ thick] (1.9,0) -- (2.6,0);
    %\draw [ thick] (3.6,-0.4) -- (3.6,-1.9);
    \draw [ thick] (2.6,-2) -- (0.4,-2);
    
    \node at (1,-0.7) {\large{$X$}};
    \draw [thick,-latex] (1.5,-0.6) -- (1.5,-1.4);
    \node at (1.5,-2.9){\normalsize{(a)}};  
%---------6-----------1.8-------------- Z
    \draw [black,thick,fill=green!30] (4.5,0) circle (0.4);
    \draw [black,thick,fill=green!30] (6,0) circle (0.4);
    \draw [black,thick,fill=green!30] (7.5,0) circle (0.4);
    \draw [black,thick,fill=green!30] (6,-2) circle (0.4);
    
    \draw [black,thick,fill=green!30] (7.5,-2) circle (0.4);
    
    %\draw [black,thick,fill=green!30] (1.8,-1.2) circle (0.4);

    \node at (4.5,0) {1};
    \node at (6,0) {2};
    \node at (7.5,0) {3};
    \node at (7.5,-2){3};
    \node at (6,-2) {2};
    \draw [ thick] (4.9,0) -- (5.6,0);
    \draw [ thick] (6.4,0) -- (7.1,0);
    %\draw [ thick] (3.6,-4.4) -- (3.6,-5.9);
    \draw [ thick] (6.4,-2) -- (7.1,-2);
    %\draw [ thick] (0,-4.4) -- (0,-5.9);
    
     \draw [thick,-latex] (6,-0.5) -- (6,-1.5);
        \node at (4.5,-.8){\large{$Z$}};

    \node at (6,-2.9){\normalsize{(b)}};

    %------------------------------------------------------------%
    % \draw [black,thick,fill=green!30] (2,-4) circle (0.4);
    % \draw [black,thick,fill=green!30] (3.8,-4) circle (0.4);
    % \draw [black,thick,fill=green!30] (5.6,-4) circle (0.4);
    % \draw [black,thick,fill=green!30] (2,-6) circle (0.4);
    
    % \draw [black,thick,fill=green!30] (5.6,-6) circle (0.4);
    
    % %\draw [black,thick,fill=green!30] (1.8,-1.2) circle (0.4);

    % \node at (2,-4) {1};
    % \node at (3.8,-4) {2};
    % \node at (5.6,-4) {3};
    % \node at (5.6,-6){3};
    % \node at (2,-6) {1};
    
    % \draw [ thick] (2.4,-4) -- (3.4,-4);
    % \draw [ thick] (4.2,-4) -- (5.2,-4);
    % %\draw [ thick] (3.6,-4.4) -- (3.6,-5.9);
    % \draw [ thick] (2.4,-6) -- (5.2,-6);
    % %\draw [ thick] (0,-4.4) -- (0,-5.9);
    
    %  \draw [thick,-latex] (3.8,-4.6) -- (3.8,-5.6);
    % \node at (3.4,-4.7) {\large {$Y$}};
    
    % \node at (3.8,-6.7){\normalsize{(c)}};

\end{tikzpicture}    
    \caption{(a) Three-qubit linear cluster, the second qubit is measured in the $X$ basis resulting in a Bell pair between qubit 1 and 3. (b) The first qubit of a three qubit linear cluster state is measured in $Z$ basis, resulting in the removal of the first qubit with a phase factor.} 
    \label{fig:2}
\end{figure}
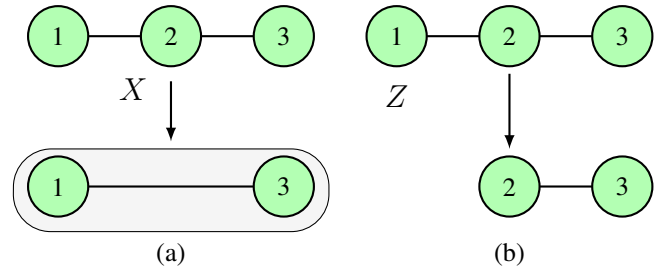
evolves as $<(-1)^{s_{2}}Z_{1}Z_{3},\ X_{1}X_{3}>$ after $X$ measurement on qubit labelled 2.
The stabilizer set are similar to the stabilizer generator of the Bell state.
Therefore, $X$ measurement results in fusion of the remaining physical qubit in cluster state into a logical qubit. In Fig. \ref{fig:2} (a) the grey shaded cluster state refers to this logical qubit.

Now if we consider the $Z$ measurement on the first qubit of the cluster state as shown in Fig. \ref{fig:2} (b), after measurement the stabilizer group transforms as $<(-1)^{s_1}X_{2}Z_{3},\ X_{3}Z_{2}>$.
However for $s_1=0$ the stabilizer group is equivalent to a two qubit cluster state and for $s_1=1$ the stabilizer group still remains same as a two qubit cluster state with a phase factor. 
Therefore we can conclude that $Z$ measurement is equivalent to cutting out a qubit from the cluster state. 

We can generalise this example to $n$ qubit linear cluster state also. 
For a cluster state consists of $\{1, 2,..., i-1,\ i,\ i+1,..., n\}$ qubits if we measure the $i^{\mathrm{th}}$ qubit in $Z$ basis then the $i^{\mathrm{th}}$ qubit will be removed from the cluster and if we measure the $i^{\mathrm{th}}$ qubit in $X$ basis the neighbouring states $\{(i-1)^{\mathrm{th}},\ (i+1)^{\mathrm{th}}\}$ will fuse together.
Thus the algorithm has offered a taste of the impact of various measurements on cluster state.
\section{Construction of non-CSS codes using measurements}
\label{sec:algorithm1}
Using Algorithm \ref{alg:Algorith1}, we now aim to build non-CSS codes by measuring qubits of the cluster state.
The steps for constructing non-CSS codes through measurement is as follows:
\begin{itemize}
    \item A set of stabilizers  $\mathcal{S'}=<S_1, S_2,\cdots, S_{n-1}>$ associated with an $[[n,1]]$ non-CSS code is given. % we shall apply our stabilizer algorithm.
    \item We construct a cluster state of $(n+1)$ qubits by strategically placing the message qubit in the cluster which we shall call as parent cluster.
    \item Let the  stabilizer generator group for $(n+1)$ qubit cluster to be $\mathcal{S}$ which will be of form Eq. \ref{eq:2}.
    \item We measure the message qubit in the $X$ basis.  
    \item Depending upon the measurement outcome, perform local unitary corrections if required. This evolves the stabilizer set from $\mathcal{S}$ to $\mathcal{S}'$. 
    %shall perform measurement the associated cluster state with the assumed initial stabilizer and shall analyse the quantum state associated with the cluster state before and after measurement. 
    %\item If the quantum state after measurement has the same set of stabilizer that we have started with we can declare our stabilizer algorithm and measurement scheme are able to construct the non-CSS code correctly.
\end{itemize}
To show the efficacy of our scheme, we consider two codes, namely $[[4,1]]$ and $[[5,1]]$ that uses four and five physical qubits respectively to encode the information of one logical qubit. 
Motivated by these examples, we generalize the technique and present it in Algorithm \ref{alg:Algorithm2}. %\textcolor{red}{Citations}
\label{sec:construction}
\subsection{Building the $[[4,1]]$ code} 
The stabilizer set for $[[4,1]]$ code is given as \cite{bell5498experimental}: 
\begin{equation}
\label{eq:3}
    \begin{split}
        \mathcal{S'}=<Y_1Z_2Z_4Y_5,\ Y_1Z_2Y_4Z_5,\ Z_1Y_2Y_4Z_5>.
    \end{split}
\end{equation}
For building the $[[4,1]]$ non-CSS stabilizer code using measurement, we reiterate the procedure given in \cite{bell5498experimental} by considering a five-qubit cluster state shown in Fig. \ref{fig:4} (a).
The quantum state associated with the five-qubit cluster state can be written as:
\vspace{-0.25cm}
\begin{equation}
    \ket{\phi}=\dfrac{1}{\sqrt{2}}\ket{0}_3\ket{+_L}+\dfrac{1}{\sqrt{2}}\ket{1}_3\ket{-_L},
    \label{eq:4}
\end{equation}
where

\begin{figure}[H]
    \centering

    \begin{tikzpicture}
    
    \draw [rounded corners =5mm,fill=gray!8] (3.8,-2.5) -- (3.8,.5) -- (7.1,0.5) -- (7.1,-2.9) -- (3.8,-2.9) -- (3.8,-2.5) ;

    % \draw [fill=black!5](3.8,-2.9) .. controls (3.65,-2) and (3.7,-1) .. (3.8,0.5) .. controls (4.8,0.6) and (5.8,0.65) .. (7.1,0.5) .. controls (7.3,-1.2) and (7.2,-2) .. (7.1,-2.9) .. controls (5.9,-3.2) and (4.8,-3.0) .. (3.8,-2.9);
    % %  (7.1,-2.9) --  (3.8,-2.9);
    
    \draw [black,thick,fill=green!30] (0,0) circle (0.4);
    \draw [black,thick,fill=green!30] (2.3,0) circle (0.4);
    \draw [black,thick,fill=green!30] (2.3,-2.3) circle (0.4);
    \draw [black,thick,fill=green!30] (0,-2.3) circle (0.4);
    \draw [black,thick,fill=green!30] (1.15,-1.15) circle (0.4);

    \node at (0,0) {2};
    \node at (2.3,0) {4};
    \node at (2.3,-2.3) {1};
    \node at (0,-2.3) {5};
    \node at (1.15,-1.15) {3};
    
    \draw [ thick] (0.4,0) -- (1.9,0);
    \draw [ thick] (2.3,-0.4) -- (2.3,-1.9);
    \draw [ thick] (0.4,-2.3) -- (1.9,-2.3);
    \draw [ thick] (0,-0.4) -- (0,-1.9);
    \draw [ thick] (0.3,-0.3) -- (0.85,-0.85);
    \draw [ thick] (2,-0.3) -- (1.45,-0.85);
    \draw [ thick] (0.3,-2.0) -- (0.85,-1.45);
    \draw [ thick] (2,-2) -- (1.45,-1.45);
    
    \node at (1.15,-0.5) {$X$};
    \node at (1.15,-1.9) {$\ket{\psi}$};
    \node at (1.15,-3.4) {(a)};
    %------------------------------------------
    
    \draw [black,thick,fill=blue!30] (4.3,0) circle (0.4);
    \draw [black,thick,fill=blue!30] (6.6,0) circle (0.4);
    \draw [black,thick,fill=blue!30] (6.6,-2.3) circle (0.4);
    \draw [black,thick,fill=blue!30] (4.3,-2.3) circle (0.4);
    %\draw [blue,thick] (3.15,-1.15) circle (0.4);

    \node at (4.3,0) {2};
    \node at (6.6,0) {4};
    \node at (6.6,-2.3) {1};
    \node at (4.3,-2.3) {5};
    %\node at (1.15,-1.15) {3};
    
    \draw [ thick] (4.7,0) -- (6.2,0);
    \draw [ thick] (4.3,-0.4) -- (4.3,-1.9);
    \draw [ thick] (4.7,-2.3) -- (6.2,-2.3);
    \draw [ thick] (6.6,-0.4) -- (6.6,-1.9);
    \node at (5.55,-1.15) {$\ket{\psi}$};
    
    \draw  [-latex](2.6,-1.15) to (3.6,-1.15);
    \node at (5.55,-3.4) {(b)};
    %\draw [fill=black!10](3.8,-2.9) -- (3.8,0.5)--(7.1,0.5) -- (7.1,-2.9) --  (3.8,-2.9);
    % \draw [ thick] (0.3,-0.3) -- (0.85,-0.85);
    % \draw [ thick] (2,-0.3) -- (1.45,-0.85);
    % \draw [ thick] (0.3,-2.0) -- (0.85,-1.45);
    % \draw [ thick] (2,-2) -- (1.45,-1.45);

\end{tikzpicture}
\begin{center}

\includegraphics[scale=0.35]{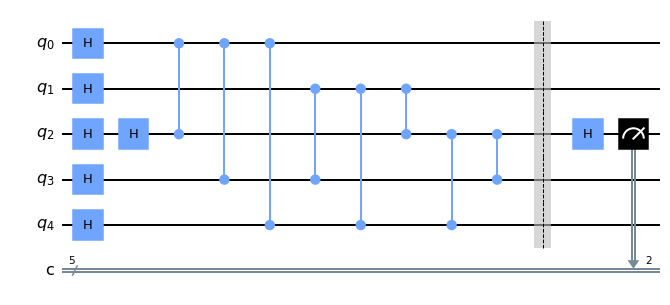}
\hspace{1cm}(c)
\end{center}

\begin{verbatim}
Stabilizers:[+ZZZIX,+ZZZXI,+IXZZZ,+XIZZZ]
Measuremnt basis: X, Input qubit:  2
New Stabilizer Set:[+ZZIX,+ZZXI,+IXZZ,+XIZZ]
Anti-commute Stabilizers:[+ZZIX,+ZZXI,
+IXZZ,+XIZZ]
Commutating Stabilizers []
Multiplication of Anti Commute Stabilizers:
[+IIXX,+ZYZY,+YZZY,+ZYYZ,+YZYZ,+XXII]
Generated set from non-commute stabilizers:
[+IIXX,+ZYZY,+YZZY]
Final set [+IIXX,+ZYZY,+YZZY]
Parity check matrix for X and Z error:
 [[0 0 0 0] [[0 0 1 1]
 [1 1 1 1]  [0 1 0 1]
 [1 1 1 1]] [1 0 0 1]]
This Code cannot correct X error:
Parity line 1 and parity line 2 is same
INDIVIDUAL,code correct Z and Y error
 \end{verbatim} 
\vspace{-1.6em}
(d)
\caption{
(a) Encoding an arbitrary state $\ket{\psi}$ into five-qubit cluster. Qubit labelled 3 is measured in the $X$ basis. (b) Measurement in $X$ basis removes the third qubit and encodes  $\ket{\psi}$ into the cluster. (c) Circuit diagram of cluster created in  QISKit  where $X$ measurement is done on the third qubit labelled as $q_2$ because numbering of qubit will start from 0. (d) Results of stabilizer set before and after the measurement using Algorithm \ref{alg:Algorith1} and generation of parity check matrix. The generated stabilizer-set has redundant stabilizers. The stabilizers of the $[[4,1]]$ can be obtained from this set.}
    
    \label{fig:3}
\end{figure}

\vspace{-0.25cm}%logical zero $\ket{0_L}$ and logical one $\ket{1_L}$ are given by 
\begin{align}
    \ket{0_L}&=\frac{1}{\sqrt{2}}\left( \ket{\Phi^-}_{15}\ket{\Phi^-}_{42}-\ket{\Psi^-}_{15}\ket{\Psi^-}_{42}\right),\\
    \ket{1_L}&=\frac{1}{\sqrt{2}} \left(\ket{\Psi^+}_{15}\ket{\Phi^+}_{42}+\ket{\Phi^+}_{15}\ket{\Psi^+}_{42}\right),\\
%\end{align}
%And
%\begin{align}
\ket{\Phi^\pm}&=\dfrac{1}{\sqrt{2}} \ket{00}\pm\dfrac{1}{\sqrt{2}}\ket{11},
\ket{\Psi^\pm}= \dfrac{1}{\sqrt{2}}\ket{01}\pm\dfrac{1}{\sqrt{2}}\ket{10}.    
\label{eq:5,6,7}
\end{align}
Using Eq. \ref{eq:2}, the stabilizers associated with the cluster state of five qubits can be written as
\vspace{-0.25cm}
\begin{equation}
\begin{split}
        \mathcal{S}= & <X_1Z_3Z_4Z_5,\ X_2Z_3Z_4Z_5,\ Z_1Z_2X_3Z_4Z_5,\\ &
        Z_1Z_2Z_3X_4,\ Z_1Z_2Z_3X_5>.
\end{split}
\label{eq:8}
\end{equation}

% \begin{equation}
% \begin{split}
%         \mathcal{S'}=<X_1Z_2Z_3X_4,\ X_2Z_3Z_4X_5,\\
%         X_1X_3Z_4Z_5,\ Z_1X_2X_4Z_5>.
% \end{split}
%     \label{eq:11}
% \end{equation}

%\end{equation}
To encode an arbitrary message, we replace the $\ket{+}$ state of the qubit at location 3 with $\alpha \ket{0}+\beta \ket{1}$ and measure it in the $X$ basis.
Due to $X$ measurement the cluster state changes to $\ket{\phi'}={X}^{s_3}({\alpha\ket{+_L}+\beta\ket{-_L}})$. % of third qubit. 
After $X$ measurement, the stabilizer generator for the modified cluster state is
\vspace{-0.25cm}
\begin{equation}
    \begin{split}
        \mathcal{S'}=<Y_1Z_2Z_4Y_5,\ Y_1Z_2Y_4Z_5,\ Z_1Y_2Y_4Z_5>,
    \end{split}
    \label{eq:9}
\end{equation}
which is equal to the stabilizer generator of the non-CSS code that we initially wanted to construct in Eq. \ref{eq:3}.
We also note that the logical $X$ and logical $Z$ operators associated with the [[4,1]] code are $\overline{X}=Z_1Z_2X_4,\ \overline{Z}=Z_1Z_2Z_4Z_5$ as verified by this construction.
The parity check matrix i.e $H = [H_x|H_z]$ \cite{2016NJPh...18j3052E} associated with Eq. \ref{eq:3} is given as :
\begin{equation}
    H=\begin{bmatrix}
\begin{array}{cccc|cccc}
     1&0&0&1 &1&1&1&1 \bigstrut[t] \\
     1&0&1&0 &1&1&1&1 \\
     0&1&1&0 &1&1&1&1 \bigstrut[b]
\end{array}\end{bmatrix}.
\label{eq:10}
\end{equation}
By considering the whole parity check matrix in Eq. \ref{eq:10}, one can show that the minimum number of linearly dependant columns are 2. Therefore the distance for this code is 2 and it can be written as $[[4,\ 1,\ 2]]$ code.
By taking $H_x$ and $H_z$ matrix individually, we can calculate the minimum distance of this code for correcting $X$ error ($d_x$) and $Z$ error ($d_z$) respectively\cite{2016NJPh...18j3052E}.
It is easy to show that two columns of $H_z$ and four columns of $H_x$ are linearly dependent.
Therefore in this code, $d_x = 2$ and $d_z = 4$.
Also, error-correction capability is verified in simulations corroborating the fact that this code corrects $Z$, $Y$ errors but detects $X$ error \cite{bell5498experimental}. 
 %\textcolor{red}{check this once more why is $I_5$ being written?}
The simulation in QISKit is done using stabilizer formation explained in \cite{aaronson2004improved}. 
We use Algorithm \ref{alg:Algorith1} to generate the stabilizer group for the $[[4,1]]$ code. 
Note that, numbering of qubits in Qiskit is started from zero.
\subsection{Building the $[[5,1]]$ code}
The stabilizer associated with the $[[5,1]]$ non-CSS code is:
\begin{equation}
\begin{split}
        \mathcal{S'}= & <X_1Z_2Z_3X_4,\ X_2Z_3Z_4X_5,\\&
        X_1X_3Z_4Z_5,\ Z_1X_2X_4Z_5>.
\end{split}
    \label{eq:11}
\end{equation}

Using the same logical flow for the construction of $[[4,1]]$ code, we have constructed a $[[5,1]]$ code with stabilizer generator $\mathcal{S'}$.

\begin{figure}[H]
    \centering
    % \includegraphics{}
    % \caption{Caption}
    % \label{fig:my_label}
    \begin{scaletikzpicturetowidth}{.5\textwidth}

    \begin{tikzpicture}[scale=\tikzscale]
    
    \draw [rounded corners=5mm,fill=gray!8] (5,0.5) -- (8.1,0.5) -- (8.1,-2.8) -- (4.2,-2.8) -- (4.2,0.5)-- (5,0.5) ;
    
    \draw [black,thick,fill=green!30] (0.2,0) circle (0.38);
    \draw [black,thick,fill=green!30] (1.7,0) circle (0.38);
    \draw [black,thick,fill=green!30] (3.2,0) circle (0.38);
    \draw [black,thick,fill=green!30] (0.2,-2.3) circle (0.38);
    
    \draw [black,thick,fill=green!30] (3.2,-2.3) circle (0.38);
    
    \draw [black,thick,fill=green!30] (1.7,-1.2) circle (0.38);

    \node at (0.2,0) {1};
    \node at (1.7,0) {2};
    \node at (3.2,0) {3};
    \node at (3.2,-2.3){4};
    \node at (0.2,-2.3) {5};
    \node at (1.7,-1.15) {6};
    
    \draw [ thick] (0.6,0) -- (1.3,0);
    \draw [ thick] (2.1,0) -- (2.8,0);
    \draw [ thick] (3.2,-0.4) -- (3.2,-1.9);
    \draw [ thick] (2.8,-2.3) -- (0.6,-2.3);
    \draw [ thick] (0.2,-0.4) -- (0.2,-1.9);
    \draw [ thick] (0.5,-0.3) -- (1.38,-0.95);
    \draw [ thick] (1.7,-0.4) -- (1.7,-0.79);
    %\draw [ thick] (1.85,-0.9) -- (2.7,-0.3);
    \draw [ thick] (2.01,-0.95) -- (2.9,-0.3);
    \draw [ thick] (2.1,-1.3) -- (2.9,-2);
    \draw [ thick] (1.3,-1.3) -- (0.45,-1.99);
    
    \draw [thick,-latex] (3.4,-1.15) -- (4.1,-1.15); 
    \node at (1.7,-1.9) {$X$};
    \node at (1.8,-3.2) {(a)};
    \node at (0.7,-1.2){$\ket{\psi}$};

    %-------------------------------------
    
    \draw [black,thick,fill=red!20] (4.7,0) circle (0.38);
    \draw [black,thick,fill=red!20] (6.2,0) circle (0.38);
    \draw [black,thick,fill=red!20] (7.6,0) circle (0.38);
    \draw [black,thick,fill=red!20] (4.7,-2.3) circle (0.38);
    
    \draw [black,thick,fill=red!20] (7.6,-2.3) circle (0.38);
    
    %\draw [black,thick,fill=green!30] (1.5,-1.2) circle (0.4);

    \node at (4.7,0) {1};
    \node at (6.2,0) {2};
    \node at (7.6,0) {3};
    \node at (7.6,-2.3){4};
    \node at (4.7,-2.3) {5};
    \node at (6.2,-1.15) {$\ket{\psi}$};
    
    \draw [ thick] (5.1,0) -- (5.8,0);
    \draw [ thick] (6.55,0) -- (7.2,0);
    \draw [ thick] (7.6,-0.4) -- (7.6,-1.9);
    \draw [ thick] (7.2,-2.3) -- (5.1,-2.3);
    \draw [ thick] (4.7,-0.4) -- (4.7,-1.9);
    
    \node at (6.6,-3.2) {(b)};
    
\end{tikzpicture}
\end{scaletikzpicturetowidth}

\begin{center}
\includegraphics[scale=0.35]{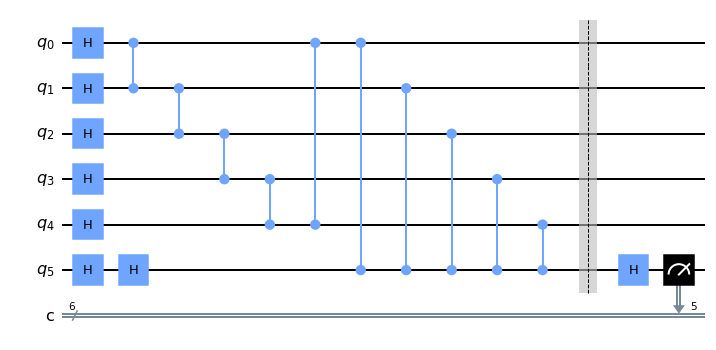}
\hspace{1cm}(c)
\end{center}
\begin{verbatim}
Stabilizers:[+ZZIIZX,+ZIIZXZ,+ZIZXZI, 
+ZZXZII, +ZXZIIZ]
Measurement basis X, Input qubit:  5
New Stabilizer:[+ZIIZX,+IIZXZ,+IZXZI,
+ZXZII,+XZIIZ]
Anti commute Stabilizers:[+ZIIZX,+IIZXZ,
+IZXZI,+ZXZII,+XZIIZ']
Commutating Stabilizers []
Multiplication of Anti Commute Stabilizers:
[+ZIZYY,+ZZXIX,+IXZZX,+YZIZY,+IZYYZ,+ZXIXZ,
+XZZXI,+ZYYZI,+XIXZZ,+YYZIZ]
Generated set from non commute stabilizers: 
[+ZIZYY,+ZZXIX,+IXZZX,+YZIZY]
Final set: [+ZIZYY,+ZZXIX,+IXZZX,+YZIZY]
Parity check matrix for X and Z error:
 [[1 0 1 1 1]   [[0 0 0 1 1]
 [1 1 0 0 0]    [0 0 1 0 1]
 [0 0 1 1 0]    [0 1 0 0 1]
 [1 1 0 1 1]]   [1 0 0 0 1]]
INDIVIDUAL,code correct X,Y and Z error
    \end{verbatim}
(d)    
    \caption{(a) Encoding an arbitrary state $\ket{\psi}$ into six-qubit cluster. Qubit labelled 6 is measured in the $X$ basis. (b) Measurement in $X$ basis removes the sixth qubit and encodes  $\ket{\psi}$ into the cluster. (c) Circuit diagram of cluster created in  QISKit  where $X$ measurement is done on the sixth qubit labelled as $q_5$ and numbering of qubit will start from 0. (d) Results of Stabilizer set before and after the measurement using Algorithm \ref{alg:Algorith1} and generation of parity check matrix. The generated stabilizer-set has redundant stabilizers. The stabilizers of the $[[5,1]]$ can be obtained from this set.
    }
    \label{fig:4}
    
\end{figure}

We consider a six-qubit cluster state as our parent cluster shown in Fig. \ref{fig:4}.
From the six-qubit cluster, we obtain the $[[5,1]]$ non-CSS code  by measuring the message qubit in the $X$ basis.
The quantum state associated with this six qubit cluster state $\ket{\phi}$ can be written as:
\vspace{-0.25cm}
%\begin{equation}
    \begin{align}
        & \mathrm{CZ}_{12}\mathrm{CZ}_{23}\mathrm{CZ}_{34}\mathrm{CZ}_{45}\mathrm{CZ}_{51}\mathrm{CZ}_{16}\mathrm{CZ}_{36}\mathrm{CZ}_{26}\mathrm{CZ}_{46}\mathrm{CZ}_{56} \ket{+}^{\otimes 6} \nonumber \\
        &= 
        % \big[ \frac{1}{\sqrt2}\ket{0}_6\{\ket{0_L}-\ket{1_L}\}+\frac{1}{\sqrt{2}}\ket{1}_6\{\ket{0_L}+\ket{1_L}\}\bigg]\\
        \dfrac{1}{\sqrt{2}}\bigg[\ket{0}_6\ket{-_L}+\ket{1}_6\ket{+_L}\bigg],
        \label{eq:12}
    \end{align}
where,
\begin{align}
    \ket{0_L}&=\dfrac{1}{4}(\ket{00000}+\ket{10010}+\ket{01001}+\ket{10100}+\ket{01010}\nonumber\\
    &\hspace{0.5cm}-\ket{11011}-\ket{00110}-\ket{11000}-\ket{11101}-\ket{00011}\nonumber \\
    &\hspace{0.5cm}-\ket{11110}-\ket{01111}-\ket{10001}-\ket{01100}-\ket{10111} \nonumber \\
    &\hspace{0.5cm}+\ket{00101}),  \\
    \ket{1_L}&=\dfrac{1}{4}(\ket{11111}+\ket{01101}+\ket{10110}+\ket{01011}+\ket{10101}\nonumber \\
    &\hspace{0.5cm}-\ket{00100}-\ket{11001}-\ket{00111}-\ket{00010}-\ket{11100}\nonumber \\
    &\hspace{0.5cm}-\ket{00001}-\ket{10000}-\ket{01110}-\ket{10011}-\ket{01000}\nonumber \\
    &\hspace{0.5cm}+\ket{11010}).
    \label{eq:13,14}
\end{align}
% $\ket{0_L}=\dfrac{1}{4}(\ket{00000}+\ket{10010}+\ket{01001}+\ket{10100}+\ket{01010}-\ket{11011}-\ket{00110}-\ket{11000}-\ket{11101}-\ket{00011}-\ket{11110}-\ket{01111}-\ket{10001}-\ket{01100}-\ket{10111}+\ket{00101})$\\
% and
% $\ket{1_L}=\dfrac{1}{4}(\ket{11111}+\ket{01101}+\ket{10110}+\ket{01011}+\ket{10101}-\ket{00100}-\ket{11001}-\ket{00111}-\ket{00010}-\ket{11100}-\ket{00001}-\ket{10000}-\ket{01110}-\ket{10011}-\ket{01000}+\ket{11010}).$\\
% \textcolor{red}{,}

The stabilizer generator for the six-qubit cluster is:
%\begin{equation}
\begin{align}
    \mathcal{S}= & <X_1Z_2Z_5Z_6,\ Z_1X_2Z_3Z_6,\ Z_2X_3Z_4Z_6,\ Z_3X_4Z_5Z_6,\nonumber\\
    & Z_1Z_4X_5Z_6,\ Z_1Z_2Z_3Z_4Z_5X_6 >.
    \label{eq:15}
\end{align}
%\end{equation}
Now instead of the $\ket{+}$ state, we use  any arbitrary state $\ket{\psi}=\alpha\ket{0}+\beta\ket{1}$ as the message to be encoded in the cluster and the encoded state in Eq. \ref{eq:12} modifies as
$\ket{\phi}= \alpha\ket{0}_6\ket{-_L}+\beta\ket{1}_6\ket{+_L}$.
Due to $X$ measurement the cluster state modifies as $\ket{\phi'}=-{X}^{s_6}({\alpha\ket{-_L}+\beta\ket{+_L}})$.
After $X$ measurement the generator of stabilizer for the modified cluster state is
\begin{equation*}
    \begin{split}
    \mathcal{S'}= & <X_1Z_2Z_3X_4,\ X_2Z_3Z_4X_5,\ X_1X_3Z_4Z_5,\ Z_1X_2X_4Z_5>.
    \end{split}
\end{equation*}
which is the desired set for the non-CSS code. 
The Parity check matrix associated with Eq. \ref{eq:11} is given as :
\begin{equation}
    H=\begin{bmatrix}
\begin{array}{ccccc|ccccc}
     1&0&0&1&0 &0&1&1&0&0 \bigstrut[t] \\
     0&1&0&0&1 &0&0&1&1&0 \\
     1&0&1&0&0 &0&0&0&1&1 \\
     0&1&0&1&0 &1&0&0&0&1  \bigstrut[b]
\end{array}\end{bmatrix}.
\label{eq:16}
\end{equation}

Also, the logical $X$ and logical $Z$ operators associated with the [[5,1]] code are \begin{align}
    \overline{X}=X_1X_2X_3X_4X_5,\ \overline{Z}=Z_1Z_2Z_3Z_4Z_5.
    \label{eq:17}
\end{align}
Similarly to $[[4,1]]$ code, one can find out $d_x = 5$ , $d_z = 5$ and  $d = 3$. Minimum distance is 3 because of there are minimum three linearly dependent columns of parity check matrix. Therefore [[5,\ 1]] code can also be written as [[5,\ 1,\ 3]]. Since $d=3$, it can correct up to $(d-1)/2$ Pauli error i.e 1 Pauli error. This result is verified by us with the help of QISKit as shown in Fig. \ref{fig:4}.

From the two examples, we can observe that constructing an $[[n,1]]$ QECC by measuring a qubit in the $(n+1)$ qubit cluster is a viable technique for non-CSS codes. We generalize this approach for any arbitrary $[[n,1]]$ non-CSS QECC by invoking the connections of the adjacency matrix of a cluster to the stabilizer group of a QECC.

\section{Construction of the cluster to encode non-CSS codes}
\label{sec:construction}
 A stabilizer code with parity check matrix $H=[H_x|H_z]$ satisfies \cite{nielsen2002quantum}
\begin{align}
  H_z H_x^T + H_xH_z^T = \mathbf{0},
  \label{eq:18}
\end{align}
where $\mathbf{0}$ denotes the null matrix.
Also the parity check matrix for a stabilizer code associated with a cluster or graph state should have the structure given below \cite{nadkarni2017recovery}:
\begin{align}
   H=[I_{n-k\times n-k}:\mathbf{0}_{n-k\times k}|A^{\mathrm{cc}}_{n-k\times n-k}:A^{\mathrm{cm}}_{n-k\times k}],
  \label{eq:19}
\end{align}
where $A^{\mathrm{cc}}_{n-k\times n-k}$ is an  adjacency matrix associated with the check nodes (qubits without message encoded on them) of an $n$ qubit cluster state, $A^{\mathrm{cm}}_{n-k\times k}$ is the adjacency matrix indicating the connection of the check nodes to the message nodes.

We propose the construction of an $(n+1)$ qubit parent cluster which upon measurement of the message qubit leads us to the desired non-CSS code. To construct an $(n+1)$-qubit cluster, we augment the matrix $H$ by adding
a row and column to both $H_x$ and $H_z$ of $H$, and call it $H'$. We solve for the unknowns in this augmented matrix using Eq. \ref{eq:18} and the fact that $H'$ should correspond to the structure of a cluster given in Eq. \ref{eq:19}.
 
% We also observe that the adjacency matrix corresponding to
% the graph state  in  Fig. \ref{fig:4} (b)
% can be constructed using the parity check matrix $H=[H_x|H_z]$ associated with the stabilizer code \cite{nadkarni2017recovery}.

\begin{figure}[H]
    \centering
    \begin{tikzpicture}
        
        \draw (0,-0.2) -- (8.8,-0.2) -- (8.8,-10.7) -- (0,-10.7) -- (0,-0.2);
        \draw (1.5,-0.7) -- (7.3,-0.7) -- (7.3,-2.3) -- (1.5,-2.3) -- (1.5,-0.7);
        \node at (4.4,-3.6)
{$\begin{bmatrix}
\begin{array}{ccccc|ccccc}
     1&0&0&1&c_0 &1&1&1&1&c_4 \bigstrut[t] \\
     1&0&1&0&c_1 &1&1&1&1&c_5 \\
     0&1&1&0&c_2 &1&1&1&1&c_6 \\
     r_0&r_1&r_2&r_3&c_3 &r_4&r_5&r_6&r_7&c_7 \bigstrut[b]
\end{array}
\end{bmatrix}$};
    \draw (1.5,-5.4) -- (7.3,-5.4) -- (7.3,-7) -- (1.5,-7) -- (1.5,-5.4);
    \node at (4.4,-8.6) {$\begin{bmatrix}
\begin{array}{cccccc|cccccc}
     1&0&0&1&0&c_0' &0&1&1&0&0&c_5' \bigstrut[t] \\
     0&1&0&0&1&c_1' &0&0&1&1&0&c_6' \\
     1&0&1&0&0&c_2' &0&0&0&1&1&c_7' \\
     0&1&0&1&0&c_3' &1&0&0&0&1&c_8'\\
     r_0'&r_1'&r_2'&r_3'&r_4'&c_4' &r_5'&r_6'&r_7'&r_8'&r_9'&c_9' \bigstrut[b]
\end{array}
\end{bmatrix}$};   
        
    \node at (4.5,-1) {$c_p$ = 0 where $p$ = 0,1,2,3,4,5,6.};    
    \node at (4.5,-1.4) {$r_q$ = 0 where $q$ = 1,5,7.};  
    \node at (4.5,-1.8) {$r_t$ = 1 where $t$ = 0,2,3,4,6.};  \node at (4.5,-5.8) {$c_i'$ = 0 where $i$ = 1,2,3,4,5,6,7,8.};
    \node at (4.5,-6.2){$r_j'$ = 0 where $j$ = 5,6,7,9.};
    \node at (4.5,-6.6){$r_k'$ = 1 where $k$ = 0,1,2,3. and $c_9'$ = 1.};
    \node at (4.5,-5){(a)};    
    \node at (4.5,-10.2){(b)};
    \end{tikzpicture}
    \caption{(a) and (b) are the parity check matrix  for 5 qubit and 6 qubit parent cluster for $[[4,1]]$ and $[[5,1]]$ QECC respectively. }
    \label{fig:5}
\end{figure}

\begin{algorithm}
\caption{Cluster formation from non-CSS stabilizer code \label{alg:Algorithm2}
i.e., forming $(n+1)$ qubit cluster from $[[n,1]]$ stabilizer code}

\begin{algorithmic}[1]
\State $\mathcal{S'}=<S_1, S_2,\cdots, S_{n-1}>=<S_{j}>_{j=1}^{n-1}$
\State Construct the $H=[H_x|H_z]$ associated with $\mathcal{S'}$.
\State Add a new row and column by inserting binary valued variables in $H_x$ and $H_z$ matrix. Call it $H'$.
\State Formulate equations of constraints for unknown variables using Eq. \ref{eq:18}.
\State Solve for the variables doing mod 2 row addition and column exchange operation on $H'$ such that it satisfies Eq. \ref{eq:19}.
\State  The matrix $H'$ associated with parent cluster state is recovered. Construct the cluster from $H'$ using the adjacency matrices $A^{\mathrm{cc}}$ and $A^{\mathrm{cm}}$. 
\State Measure the message qubit in $X$ basis and depending upon the measurement outcome, perform local unitary corrections if required to get $\mathcal{S'}$ containing $(n-1)$ stabilizer generators of the desired $[[n,1]]$ non-CSS code.
\end{algorithmic}
\end{algorithm}

\vspace{-0.3cm}

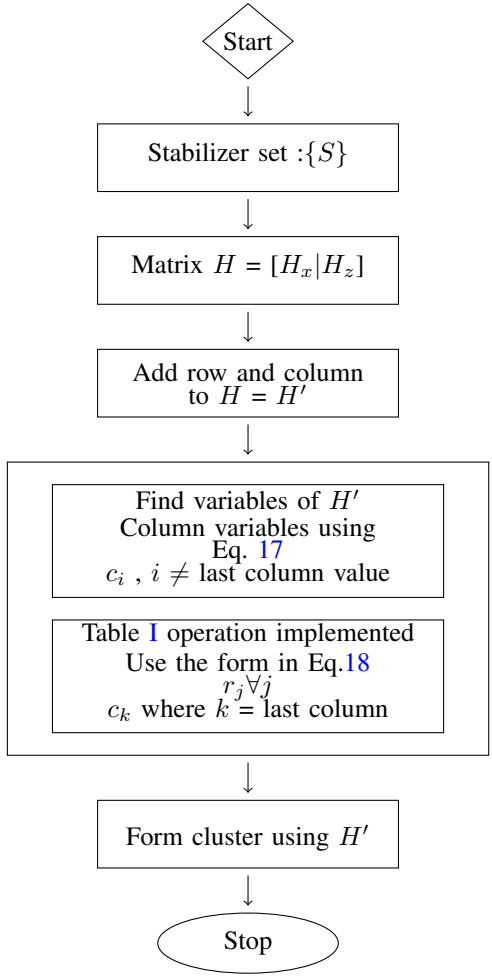
\begin{figure}[H]
    \centering
    
    \begin{tikzpicture}
        
        \draw (0,0) -- (0.6,-0.5) -- (0,-1) -- (-0.6,-0.5) -- (0,0);
        
        \draw [->] (0,-1.1) -- (0,-1.5);
        
        \draw (0,-1.6) -- (2,-1.6) -- (2,-2.5) -- (-2,-2.5) -- (-2,-1.6) -- (0,-1.6);
        
        \draw [->] (0,-2.6) -- (0,-3);
        
        \draw (0,-3.1) -- (2,-3.1) -- (2,-4) -- (-2,-4) -- (-2,-3.1) -- (0,-3.1);
        
        \draw [->] (0,-4.1) -- (0,-4.5);
        
        \draw (0,-4.6) -- (2,-4.6) -- (2,-5.5) -- (-2,-5.5) -- (-2,-4.6) -- (0,-4.6);
        
        \draw [->] (0,-5.6) -- (0,-6);
        
        \draw (0,-6.1) -- (3.2,-6.1) -- (3.2,-10) -- (-3.2,-10) -- (-3.2,-6.1) -- (0,-6.1);
        
        \draw (0,-6.4) -- (2.6,-6.4) -- (2.6,-7.9) -- (-2.6,-7.9) -- (-2.6,-6.4) -- (0,-6.4);
        
        \draw (0,-8.2) -- (2.6,-8.2) -- (2.6,-9.7) -- (-2.6,-9.7) -- (-2.6,-8.2) -- (0,-8.2);
        
        \draw [->] (0,-10.1) -- (0,-10.5);

        \draw (0,-10.6) -- (2,-10.6) -- (2,-11.5) -- (-2,-11.5) -- (-2,-10.6) -- (0,-10.6);
        
        \draw [->] (0,-11.6) -- (0,-12);

        \draw (0,-12.5) ellipse (1.2cm and 0.4cm);
        
        \node at (0,-12.5) {Stop};
        \node at (0,-11.1) {Form cluster using $H'$};
        
        \node at (0,-8.4) {Table \ref{tab:R and C operation} operation implemented } ;
        \node at (0,-8.8) {Use the form in Eq.\ref{eq:18}};
        \node at (0,-9.1){$r_j \forall j$};
        
        \node at (0,-9.4){$c_k$ where $k$ = last column};
        
        %%% text %%%%
        
        \node at (0,-0.5) {Start};
        \node at (0,-2) {Stabilizer set :\{$S$\}};
        \node at (0,-3.5) {Matrix $H$ = [$H_x|H_z$]};
        
        \node at (0,-4.9) {Add row and column};
        \node at (0,-5.2) {to $H$ = $H'$};
        \node at (0,-6.6) {Find variables of $H'$};
        \node at (0,-7){Column variables using };
        \node at (0,-7.3){Eq. \ref{eq:17}}; 
        \node at (0,-7.6){$c_i$ , $i$ $\neq$ last column value};

        %\node at (0,-6.8) {using eq.\ref{eq:13} and eq.\ref{122} };
       
    \end{tikzpicture}
    
    \caption{Flow diagram of Algorithm 2}
    \label{fig:my_label}
\end{figure}

\newpage

To this end, we perform row operations on $H'$ giving us the required $n+1$-qubit cluster and its $A^{\mathrm{cc}}$ and $A^{\mathrm{cm}}.$ 
% In this context of parity check matrix, one should consider only modulo 2 row addition and column exchange operations.
In this context of parity check matrix, one should consider the operations as explained in Table. \ref{tab:R and C operation}.
\vspace{-0.1cm}
\begin{table}[H]
    \centering
    \begin{tabular}{|c||c|}
      \hline 
      
    OPERATIONS & EQUIVALENCE  \\ \hline \hline
   Modulo 2 row operation & Multiplying two stabilizers of the set. \\ \hline
    Column exchange operations &   Relabelling of qubits. \\ \hline 
    \end{tabular}
    \caption{}
    \label{tab:R and C operation}

\end{table}
% While modulo 2 row addition is equivalent of multiplying two stabilizers of the set, a column exchange operation denotes relabelling of qubits.

\vspace{-0.5cm}
For example the parity check matrix associated with the $[[4,1]]$ code and $[[5,1]]$ code are given in Eq. \ref{eq:10} and Eq. \ref{eq:16}.

The augmented parity check matrix  $H'$ that satisfies Eq. \ref{eq:18} and Eq. \ref{eq:19} are given in Fig \ref{fig:5}. With the help of row additions and column exchange, we solve unknown variables and obtain the 5-qubit cluster and 6-qubit cluster corresponding to $[[4,1]]$ and $[[5,1]]$ code respectively.  As a result we can  we construct the $(n+1)$ qubit parent cluster and measure the message qubit in the $X$ basis giving an $[[n,1]]$ non-CSS code.

We summarize this approach in Algorithm \ref{alg:Algorithm2} which builds $(n+1)$ qubit cluster given $(n-1)$ stabilizers of the $[[n,1]]$ non-CSS code.

\vspace{-0.3cm}

\section{Conclusion}
\label{sec:conclude}
In this paper we used cluster states to design non-CSS codes by clever single-qubit measurements. 
We showed the evolution of stabilizers under single-qubit measurements and the impact of measurements on cluster state. If we are given non-CSS code stabilizers with parameters $[[n,1]]$, we consider $(n+1)$ qubit cluster with one message qubit, that is strategically located in this cluster. By measuring the message qubit in the $X$ basis, followed by appropriate local operations, we project the unmeasured qubits into the given non-CSS code.  
To summerise, any non-CSS stabilizer code can be constructed from an $(n+1)$-qubit cluster state. 
\vspace{-0.1cm}
\section*{Acknowledgment}
A. R. thanks the start-up grant from IISER Bhopal.
S.S acknowledges funding support for Chanakya -PG fellowship from the National Mission on Interdisciplinary Cyber Physical Systems, of the Department of Science and Technology, Govt. of India through the I-HUB Quantum Technology Foundation.
% The authors thank everyone who helped in the work of this paper. % A. R. thanks the start-up grant from IISER Bhopal. S. S. thanks the Chanakya PG fellowship from i-HUB, QTF IISER Pune. 

%\textcolor{red}{Bibliography should be in IEEE format. Please go through the format}
%\vspace{-0.5cm}
\bibliographystyle{IEEEtran} 
\bibliography{mybiblography.bib}

\end{document}